# Spin-dependent transport properties of a GaMnAs-based vertical spin metal-oxide-semiconductor field-effect transistor structure

Toshiki Kanaki,[1,a)] Hirokatsu Asahara,[1] Shinobu Ohya,[1,b)] and Masaaki Tanaka[1,c)]

[1]*Department of Electrical Engineering and Information Systems, The University of Tokyo, 7-3-1 Hongo, Bunkyoku, Tokyo 113-8656, Japan*

We fabricate a vertical spin metal-oxide-semiconductor field-effect transistor (spin-MOSFET) structure, which is composed of an epitaxial single-crystal heterostructure with a ferromagnetic-semiconductor GaMnAs source/drain, and investigate its spin-dependent transport properties. We modulate the drain-source current $I_{DS}$ by ~±0.5 % with a gate-source voltage of ±10.8 V and also modulate $I_{DS}$ by up to 60 % with changing the magnetization configuration of the GaMnAs source/drain at 3.5 K. The magnetoresistance ratio is more than two orders of magnitude higher than that obtained in the previous studies on spin MOSFETs. Our result shows that a vertical structure is one of the hopeful candidates for spin MOSFET when the device size is reduced to a sub-micron or nanometer scale.

[a)]Electronic mail: kanaki@cryst.t.u-tokyo.ac.jp
[b)]Electronic mail: ohya@cryst.t.u-tokyo.ac.jp
[c)]Electronic mail: masaaki@ee.t.u-tokyo.ac.jp

One of the most crucial issues in existing Si-based semiconductor electronics is to develop devices for the post-scaling era. Spintronics devices, which utilize not only charge degrees of freedom but also spin degrees of freedom, have received considerable attention. Above all, spin metal-oxide semiconductor field-effect transistors (spin MOSFETs[1,2]), whose source and drain are ferromagnetic, are one of the most promising devices because of their compatibility with sophisticated semiconductor technologies and their variety of applications such as reconfigurable logic circuits[3] and non-volatile power gating.[4] In previous studies on spin MOSFETs,[5,6] the drain-source current was controlled by the gate-source voltage and magnetization configuration of the source and drain; however, the magnetoresistance (MR) ratios (0.1%[5] and 0.005%[6]) were too small to be put into practical applications in these studies, and thus spin MOSFET with a high MR ratio is strongly required.

In this study, we study a vertical spin-MOSFET structure, in which the current flows perpendicular to the plane and is subject to the gate-electric field from the side of the mesa (Fig. 1(a)), while in conventional planar device structures the current flows parallel to the plane and is subject to the gate-electric field from the top. From the viewpoint of performance as a MOSFET, the vertical structure is preferable to the planar structure because of steep switching behavior and reduction of the circuit area.[7] The vertical structure is also preferable for the spin-dependent transport because the channel length can be easily reduced and controlled with an atomic level, which is suitable for higher MR ratios because we can minimize the influence of spin relaxation during the transport of carriers in the channel. In our study, we use ferromagnetic semiconductor GaMnAs as a source/drain material because high-quality single-crystal GaMnAs can be epitaxially grown with atomically flat interfaces with various semiconductors (GaAs,[8] AlAs,[9] InGaAs,[10] and ZnSe[11]). Actually so far, high MR ratios up to 290 % (at 0.39 K[8]) have been reported in GaMnAs-based single-barrier structures.

Figure 1(a) shows a schematic illustration of the GaMnAs-based vertical spin-MOSFET structure investigated in this study, which comprises $Ga_{0.93}Mn_{0.07}As$ (20 nm)/GaAs (9 nm)/$Ga_{0.95}Mn_{0.05}As$ (20 nm)/GaAs:Be (50 nm), from the top to the bottom, grown on a $p^+$-GaAs (001) substrate by low-temperature molecular beam epitaxy. The substrate temperatures during the growth of the $Ga_{0.93}Mn_{0.07}As$, GaAs, $Ga_{0.95}Mn_{0.05}As$, and GaAs:Be layers were 205 °C, 205 °C, 225 °C, and 540 °C, respectively. The device fabrication process is as follows. Columnar mesas with 200 μm in diameter were formed by standard photolithography and chemical wet etching. Etching was stopped so that the remaining bottom $Ga_{0.95}Mn_{0.05}As$ was reduced to 10 nm in thickness. Then, 27-nm-thick $AlO_x$ was deposited by atomic layer deposition at a substrate temperature of 150 °C. After contact holes were opened by buffered fluoric acid on the top of the mesas, we deposited Au and isolated the drain and gate electrodes by photolithography and chemical wet etching. As shown in Fig. 1(a), the upper (lower) GaMnAs layer is the drain (source), the intermediate GaAs layer is the channel, and the sidewall $AlO_x$/Au is the gate. Figure 1(b) shows a top-view photograph of the fabricated GaMnAs-based vertical spin-MOSFET structure. Figures 1(c) and 1(d) show the expected band diagrams at the side surface of the GaMnAs/GaAs/GaMnAs columnar mesa when a drain-source voltage $V_{DS}$ is applied. It has been reported that the potential height of GaAs for holes in GaMnAs is approximately 100 meV.[12] Note that holes in GaMnAs are in the impurity band (IB) in the band gap and the Fermi level $E_F$ is located in the IB.[13,14,15] Since the position of $E_F$ depends on the Mn content $x$ in $Ga_{1-x}Mn_xAs$,[13,14] we estimate $E_V - E_F$ to be ~ 40 meV (in terms of hole energy) in our device, where $E_V$ and $E_F$ are the valence band edge and Fermi level, respectively. When a gate-source voltage $V_{GS}$ is not applied (Fig. 1(c)), the drain-source current $I_{DS}$ is suppressed due to the potential height of GaAs for holes. When a negative $V_{GS}$ is applied (Fig. 1(d)), the $I_{DS}$ can flow due to the band bending in the GaAs layer. The gate-electric field modulates the current ($I_{DS}$) flowing perpendicular to the plane in the region of a few nm from the side surface of the GaMnAs/GaAs/GaMnAs mesa.

Figure 2(a) shows the $I_{DS} - V_{DS}$ characteristics measured at 3.5 K in parallel magnetization configuration. It has been reported that low-temperature grown GaAs works as a barrier for holes in GaMnAs when its thickness is larger than 6 nm.[8,16,17] The resistance area (*RA*) product at $V_{DS}$ of 1 mV was 0.58 $\Omega$cm$^2$, which agrees with the value obtained in the previous work,[17] meaning that the electrical property of the fabricated magnetic tunnel junction in this study is similar to the one obtained in the previous work. Figure 2(b) shows *RA* as a function of the magnetic field $\mu_0 H$ applied along the in-plane easy magnetization axis of GaMnAs (at a 10-degree angle from the [100] direction toward the [1$\bar{1}$0] direction) when $V_{DS}$ is 1 mV. A clear tunneling magnetoresistance (TMR) was observed and the TMR ratio, which is defined as [*RA*(*H*)–*RA*(0)]/*RA*(0)×100, shown in the right axis amounted to 60 % in antiparallel magnetization configuration at 3.5 K. Here, *RA*(*H*) is the *RA* value with a magnetic field *H* and *RA*(0) is 0.56 $\Omega$cm$^2$ in the parallel magnetization configuration in the major loop (black solid curves). We observed a clear minor loop (a red dotted curve in Fig. 2(b)), indicating that the antiparallel configuration is stable. The coercive force obtained by sweeping the magnetic field from negative to positive in the minor loop is different from the value obtained in the major loop because the coercive force depends on the previous magnetization process. Small terraces shown in the major loop at around ±0.015 T and ±0.02 T correspond to the 90° magnetization configuration of the GaMnAs layers. In Fig. 2(c), the black solid lines express the $\Delta I_{DS}(V_{GS}) - V_{GS}$ characteristics with various $V_{DS}$ in parallel magnetization configuration at 3.5 K, where $\Delta I_{DS}(V_{GS})$ is defined as $I_{DS}(V_{GS}) - I_{DS}(V_{GS} = 0 \text{ V})$. The value of $V_{GS}$ was changed with a step of 1.35 V. There is a clear tendency that $|\Delta I_{DS}|$ increases with increasing $-V_{GS}$ when $V_{GS} < 0$, which means that $I_{DS}$ is effectively modulated by $V_{GS}$. Note that the sign of $\Delta I_{DS}(V_{GS})$ is reversed with respect to $V_{DS} = 0$ V because the sign of $I_{DS}$ is reversed. While $\Delta I_{DS}$ is changed by $V_{GS}$ at $V_{GS} < 0$, $\Delta I_{DS}$ is insensitive to $V_{GS}$ at $V_{GS} >$ 0, which is probably caused by the interface states at the AlO$_x$/GaAs interface. The modulation of $I_{DS}$ at $V_{GS}$ = 0 ~ −10 V was up to ~±0.5 %. The inset in Fig. 2(c) shows the $V_{GS}$ dependence of $\Delta I_{DS}(V_{GS})$ when $V_{DS}$ = −0.1 V. The maximum value of $|\Delta I_{DS}(V_{GS})|$ is ~ 1 μA. The leakage current was at most a few nA, and thus it is negligible. We note that we obtained a larger modulation of $I_{DS}$ up to ~ ±1 % for $V_{GS}$ = ±7 V and a much lower leakage current of tens of pA in another similar GaMnAs-based vertical spin-MOSFET device structure. Figure 3 shows (−$I_{DS}$) − (−$V_{DS}$) characteristics at around $V_{DS}$ = −100 mV with $V_{GS}$ of ±10.8 V in the parallel and antiparallel magnetization configurations. The red (blue) solid line was obtained at $V_{GS}$ = 10.8 V in the parallel (antiparallel) magnetization configuration and the green (black) dotted line was obtained at $V_{GS}$ = −10.8 V in the parallel (antiparallel) magnetization configuration. We see that $I_{DS}$ is controlled not only by $V_{GS}$ but also by the magnetization configuration.

Figure 4(a) shows the TMR ratios as a function of $V_{GS}$ when $V_{DS}$ = −70 mV and 8 mV. The TMR ratios increased by several % (shown at the right axis) with increasing $V_{GS}$. In Fig. 4(b) the black solid lines express the TMR ratio normalized by that at $V_{GS}$ = 0 V, as a function of $V_{GS}$ with various $V_{DS}$. The TMR ratio tends to increase with increasing $V_{GS}$ for all $V_{DS}$.

Here, we discuss the possible origins of the TMR modulation by $V_{GS}$. In the GaAs channel, the application of $V_{GS}$ induces the change in the potential height for holes. According to the theoretical calculation on the potential-height dependence of the TMR ratio,[18] while the TMR ratio becomes larger when the potential height becomes larger (corresponding to the case of $V_{GS}$ > 0 V), the TMR ratio becomes smaller when the potential height becomes smaller (corresponding to the case of $V_{GS}$ < 0 V). This scenario agrees with our experimental result. Also, we discuss the effect of the parasitic resistance composed of the top/bottom GaMnAs, GaAs:Be, and $p^+$-GaAs substrate. The equivalent circuit of our device can be expressed as a series connection of the resistance of the magnetic tunnel junction and the parasitic resistance composed of the top/bottom GaMnAs, GaAs:Be, and $p^+$-GaAs(001) substrate. Because the bottom GaMnAs is located beneath the gate electrode, it would be subject to the large gate electric field. Thus, the voltage applied to the magnetic tunnel junction would change with $V_{GS}$ due to the change of the parasitic resistance, which could lead to the change of $I_{DS}$ and the TMR ratio. However, we found that this does not happen for the following reason: From the GaAs-barrier thickness dependence of $RA$ of GaMnAs/GaAs/GaMnAs magnetic tunnel junctions obtained in the previous study,[17] the parasitic resistance is estimated to be $8\times10^{-5}$ Ω. This value of the parasitic resistance is negligibly small in comparison with the experimental modulation of the resistance (at least 0.1 Ω obtained when $V_{DS}$ was –0.1 V) in our spin-MOSFET device. Then, we discuss the change of the effective cross section of the columnar mesa induced by the change of the depletion layer in the GaMnAs layer. If the change of $I_{DS}$ at $V_{DS}$ = –0.1 V due to the gate electric field were attributed only to the change of the effective cross section of the columnar mesa due to the change of the surface depletion layer thickness of GaMnAs, the change of the radius would be 61 nm, which is too large in comparison with the change of the depletion layer thickness of GaMnAs (only a few nm) when the gate electric field is applied to it.[19] Thus, this scenario is unrealistic. For the reasons described above, we conclude that the intrinsic gate-electric-field effect on the GaAs channel layer is responsible for the change of $I_{DS}$ observed in our experiment.

In summary, we fabricated the GaMnAs-based vertical spin-MOSFET structure and investigated its spin-dependent properties. We controlled $I_{DS}$ not only by $V_{GS}$ but also by the magnetization configuration of the ferromagnetic GaMnAs source and drain. The TMR ratio amounted to 60%, which is much larger than that obtained in the previous studies on the planar spin MOSFETs.[5,6] The TMR ratio was also modulated by $V_{GS}$. We found that the parasitic resistance is negligible and that the gate-electric-field effect on the intermediate

GaAs layer is mainly responsible for the change of $I_{DS}$. Our result demonstrates the possibility that our device can provide desired properties of spin MOSFET when the device size is reduced to a sub-micron or nanometer scale.

The authors thank Prof. S. Takagi, Prof. M. Takenaka, Prof. D. Chiba, and Dr. Koyama for valuable discussions and technical helps. This work is supported by Grants-in-Aid for Scientific Research including the Specially Promoted Research and the Project for Developing Innovation Systems of MEXT. Part of this work was carried out under the Cooperative Research Project Program of RIEC, Tohoku University. T. K. acknowledges financial support from JSPS through the Program for Leading Graduate Schools (MERIT).

Figure captions

FIG. 1. (Color online) (a) Schematic cross-sectional structure of the GaMnAs-based vertical spin MOSFET investigated in this study, which is composed of $Ga_{0.93}Mn_{0.07}As$ (20 nm)/GaAs (9 nm)/$Ga_{0.95}Mn_{0.05}As$ (20 nm)/GaAs:Be (50 nm) grown on a $p^+$-GaAs (001) substrate. The upper (lower) GaMnAs layer is the drain (source), the intermediate GaAs layer is the channel, and the sidewall $AlO_x$/Au is the gate. The drain-source voltage, gate-source voltage, and drain-source current are defined as $V_{DS}$, $V_{GS}$, and $I_{DS}$ (represented by an orange arrow), respectively. (b) Top-view photograph of the fabricated vertical spin-MOSFET structure. (c)(d) Expected band diagram at the side surface of the GaMnAs/GaAs/GaMnAs columnar mesas when $V_{GS} = 0$ V (c) and when $V_{GS} < 0$ V (d), where $e$, $E_v$, and $E_F$ are the elementary charge, top of the valence band, and Fermi level, respectively.

FIG. 2. (Color online) (a) Drain-source current ($I_{DS}$) as a function of the drain-source voltage ($V_{DS}$) at 3.5 K. (b) $RA$ (left axis) and the TMR ratio (right axis) as a function of the in-plane magnetic field $\mu_0 H$ applied at a 10-degree angle from the [100] direction toward the [$\bar{1}10$] direction at 3.5 K. The black solid line corresponds to a major loop and the red dotted line to a minor loop. The red arrows show the magnetic field sweep direction. The TMR ratio amounted to 60 %. The magnetization direction in the major loop is indicated by black arrows. (c) The black solid lines represent the $\Delta I_{DS}(V_{GS}) - V_{GS}$ characteristics for various $V_{DS}$ in the parallel magnetization configuration at 3.5 K and 0 T, where $\Delta I_{DS}(V_{GS})$ is defined as $I_{DS}(V_{GS}) - I_{DS}(V_{GS} = 0$ V$)$. We changed $V_{GS}$ with a step of 1.35 V. The inset shows the $V_{GS}$ dependence of $\Delta I_{DS}(V_{GS})$ when $V_{DS} = -0.1$ V.

FIG. 3. (Color online) Drain-source current ($-I_{DS}$) as a function of the drain-source voltage ($-V_{DS}$) at 3.5 K with a gate-source voltage $V_{GS}$ of $\pm 10.8$ V in the parallel and antiparallel magnetization configurations at around $V_{DS} = -100$ mV. The red (blue) solid line was obtained at $V_{GS} = 10.8$ V in the parallel (antiparallel) magnetization configurations and the green (black) dotted line was obtained at $V_{GS} = -10.8$ V in the parallel (antiparallel) magnetization configurations. The data in the antiparallel magnetization configuration was measured at 0 T in the minor loop.

FIG. 4. (Color online) (a) TMR ratio (left axis) and normalized TMR ratio by that at $V_{GS} = 0$ V (right axis) as a function of $V_{GS}$ when $V_{DS} = -70$ mV and 8 mV. The red lines are the guide to the eye. (b) TMR ratio normalized by that at $V_{GS} = 0$ (the solid black lines) as a function of $V_{GS}$ for various $V_{DS}$. The data in the antiparallel magnetization configuration was measured at 0 T in the minor loop. All data were obtained at 3.5 K.

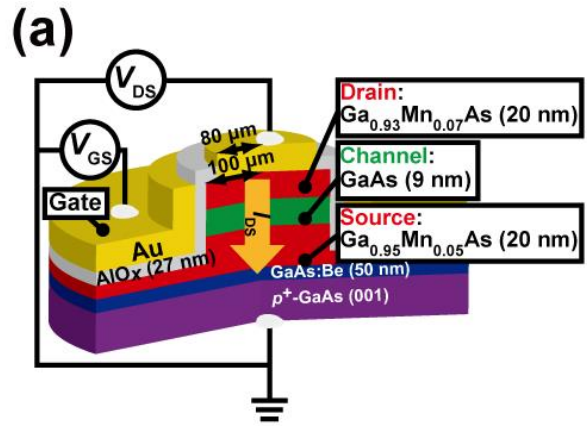

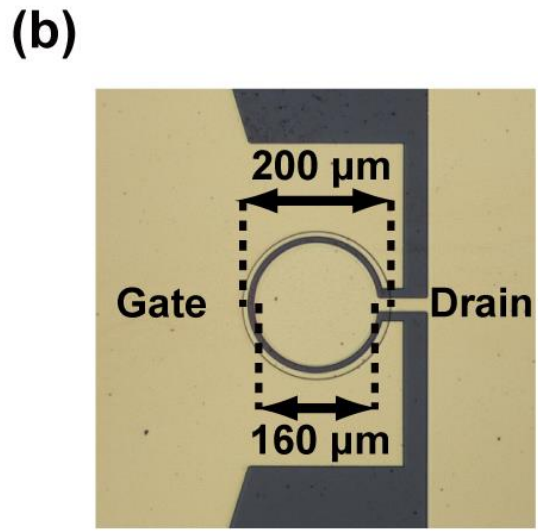

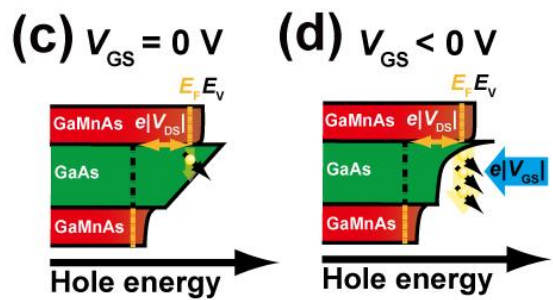

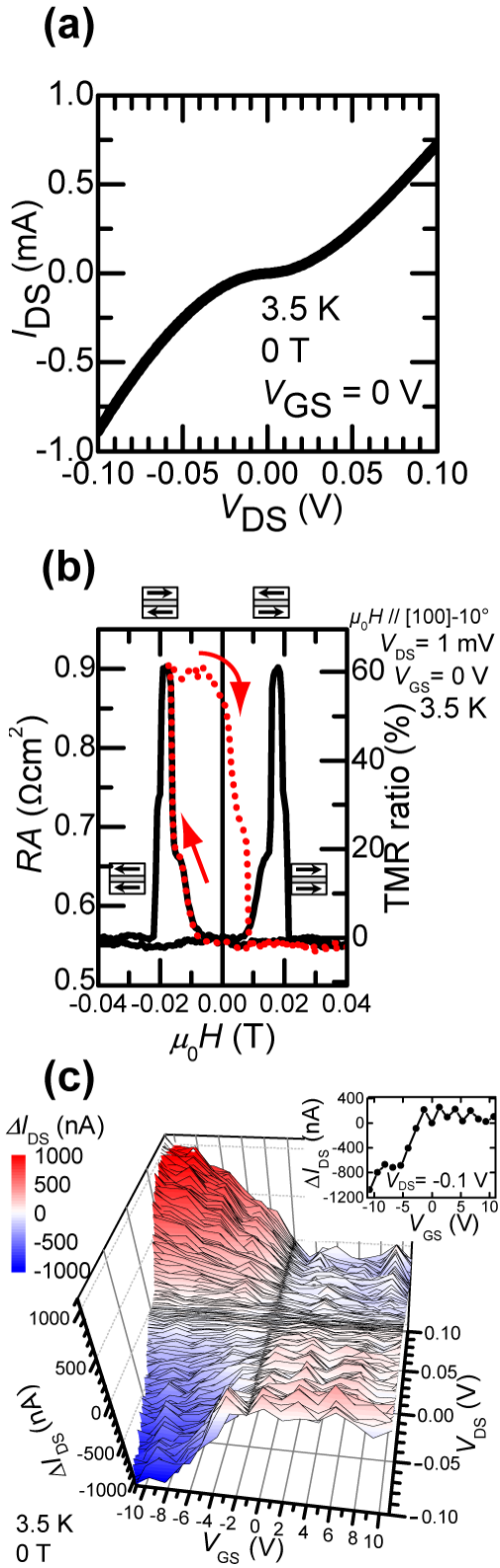

Fig. 2

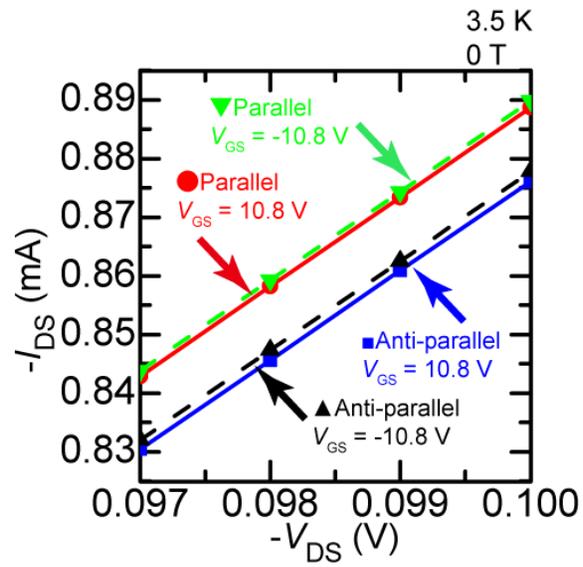

Fig. 3

**(a)**

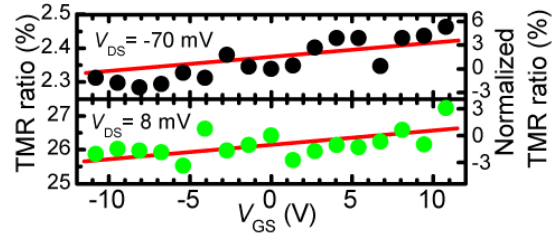

**(b)**

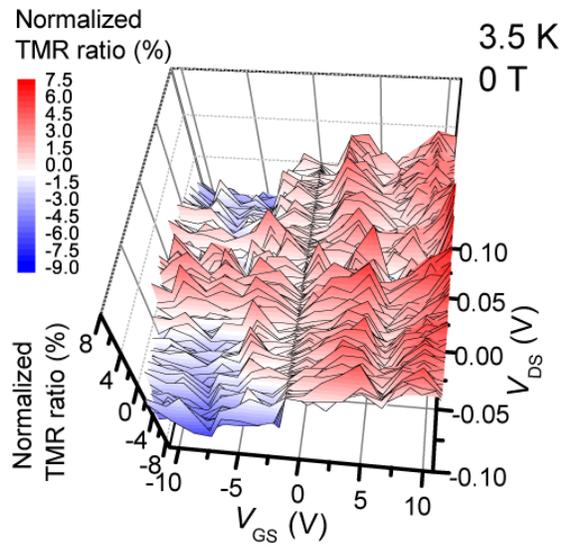

Fig. 4